\documentstyle[12pt]{article}
\begin{document}
\topmargin 0pt
\oddsidemargin 5mm
\headheight 0pt
\headsep 0pt
\topskip 9mm

\hfill UT-Komaba-93-04

\hfill March  1993
\begin{center}

{\large \bf

Exponent of n-Ising matter fields coupled to $2d$ gravity }

\vspace{24pt}

{\sl Shinobu Hikami}

\vspace{6pt}

Department of Pure and Applied Sciences, University\\
of Tokyo, Meguro-ku, Komaba 3-8-1, Tokyo 153, Japan\\
and\\
Laboratoire de Physique Th\'eorique de l$^{\prime}$Ecole
Normale Sup\'erieure,\\
 24 rue Lhomond,F-75231 Paris Cedex 05,
France\\
\end{center}

\vspace{24pt}

\begin{center}

{\bf Abstract}
\end{center}
\vspace{6pt}
   n - Ising spins
 on a random
surface  represented by a matrix model is studied as a model of
the 2D gravity coupled to matter field with the central charge
$c > 1$.
The magnetic field is introduced to discuss the scaling exponent $\Delta$
, and the value of this  magnetic field
exponent is estimated by the series expansion.

\newpage
\addtolength{\baselineskip}{0.20\baselineskip}

  The behavior of 2d gravity coupled to  a matter field
has been studied by a matrix model in the large N limit [1].
The analytic solution by Liouville theory is  consistent with the
result of the matrix model [2]. However, these exact results are all
restricted to  the case of the central charge of the matter
 field $c\leq 1$, and
the correct expression for the string susceptibility $\gamma_{st}$ for
the $c > 1$ remains unsolved.

 As a matrix model for 2d gravity coupled to matter field of $c > 1$,
a multiple n-Ising spin model has been introduced. The series
expansion [3,4]
and  Monte Carlo simulations [5,6,7] have been performed
to study the universality of the phase transition. These
results suggest the existence
of the intermediate region for $c > 1$, where
the phase transition remains the
second order with nontrivial critical exponents.

 In this letter, we extend the perturbational analysis
of the previous studies [3,4]
by adding the external magnetic field
on the  Ising spin, and discuss the scaling relation of this
multiple n-Ising model coupled to 2d gravity.

 The free energy of the one-Ising model coupled to 2d gravity
in the presence of the  external magnetic field $H$ [8]  is given by
\begin{eqnarray}
 F & = &  - ln Z \nonumber \\
   & = &  - ln \int dM_{+}dM_{-} exp [
 - tr({1\over{2}} M_{+}^{2}+
   {1\over{2}}M_{-}^{2} + aM_{+}M_{-} - g e^{H}M_{+}^{4} \\
   &   &  - g e^{-H} M_{-}^{4}) ] \nonumber
\end{eqnarray}

\noindent
where $M_{+}$ and $M_{-}$ are Hermitian $ N\times N$ matices and
they represent the vertices on which the up and down Ising spins
are placed respectively . The quantity $a$ is
a coupling constant related to the temperature
 $\beta = {1/ kT}$ as
\begin{equation}
 a = \exp{(-2\beta)}
\end{equation}

In the large $N$ limit, the free energy is dominated
 by the planar diagrams and
it has a scaling form  near the critical cosmological constant $g_{c}$,
\begin{equation}
 F = ( g - g_{c} )^{2-\gamma_{st}}\psi( {H\over{(g - g_{c})^{\Delta}}} )
\end{equation}
 We restrict ourselves to the genus zero topology.
  The coupling constant $a$ has a critical value $a_{c}$ at which
 new universal
behavior appears.  For the one-Ising model,  the critical value
$a_{c}$ is 1/4, and the string susceptibility exponent $\gamma_{st}$ changes
from -1/2 of the pure gravity case$(a\not= 1/4)$ to -1/3 of the Ising case
at $a=1/4$. The exponent $\gamma_{st}$ and the magnetic field exponent
 $\Delta$ are known for the unitary (p,q)
model [9] in which the central charge $c$ is given by
$c = 1 - 6(p-q)^{2}/pq $  for $c < 1$;

\begin{equation}
 \gamma_{st} = { - {2\over {p + q -1 }}}
\end{equation}
\begin{equation}
 \Delta = ({p\over 2}+ q)/( p + q - 1 )
\end{equation}

The Ising case  corresponds to p=4 and q=3 (c=1/2),
 and thus we have $ \gamma_{st}= -1/3$, $\Delta = 5/6$.
There exists a relation between these two exponents,
 $\Delta=(3-\gamma_{st})/4$ for $c\leq 1$.

 We extend the one-Ising model to the multiple n-Ising model, which
is described now  by $2^{n}$ matrices [3]. Each Ising spin of n-species
has up and down two states, and there are $2^{n}$ different spin
states on
the vertices of the diagrams.
 Each Ising spin has
the central charge 1/2 and total central charge is simply n/2,
which becomes a central charge of the matter field $c=n/2$.

For the n=2 (c=1) case, the magnetic field  exponent
 $\Delta$ may be obtained from the
general expression of the (p,q) model
in the limit of $p=q+1$, $p\rightarrow \infty$ as $\Delta = 3/4$.
We will show that indeed this value is obtained by our series expansion.

 We use the series expansion about the cosmological constant $g$ for the
fixed topology(genus zero), from which the exponents $\gamma_{st}$ and $\Delta$
are
extracted by the ratio method. We have already obtained the string
susceptibility
exponent $\gamma_{st}$ by the use of
the series expansion up to order eight [4]
. The coefficients of this series of the m-th derivative of the
free energy behaves as

\begin{equation}
 \lim_{H\rightarrow 0} d^{m}F/d^{m}H = \sum c_{k}g^{k}
\end{equation}

\begin{equation}
 c_{k} \sim A^{k}k^{-3 + \gamma_{st} + m\Delta }
\end{equation}

 The magnetic field $H$ is applied to one spin component and (n-1) spins remain
free of this external  magnetic field. The free energy $F$ under this magnetic
field
is easily calculated in each Feynman diagram.  We develop the series
expansion up to order eighth including the magnetic field.
For example, up to order $g^{3}$, the free energy of n-
Ising model with a magnetic field $H$ becomes,

\begin{eqnarray}
 F & = & -2(e^{H}+e^{-H})g + [16( e^{2H}+e^{-2H} + 2a^{2})
(1+a^{2})^{n-1} \nonumber \\
   & + & 2(e^{2H}+e^{-2H}+2a^{4})(1+a^{4})^{n-1}]g^{2} \nonumber \\
  & -  & \{128[e^{3H}+e^{-3H}+(e^{H}+e^{-H})(2a^{2}+a^{4})]
(1+2a^{2}+a^{4})^{n-1} (8)\nonumber \\
  & +  & {256\over {3}}[e^{3H}+e^{-3H}+(e^{H}+e^{-H})
(3a^{2})](1+3a^{2})^{n-1} \\
  & +  &
64[e^{3H}+e^{-3H}+(e^{H}+e^{-H})(a^{2}+2a^{4})](1+a^{2}+2a^{4})^{n-1}\nonumber
\\
  & +  &  {32\over{3}}[e^{3H}+e^{-3H}+(e^{H}+e^{-H})(3a^{4})](1+3a^{4})^{n-1}
\}g^{3} \nonumber \\
  & +  & \dots \nonumber
\end{eqnarray}

\noindent
where the factor $(1-a^{2})^{-2}$ is absorbed in the redefinition of
$g$.

The quantity $A$ is the inverse of the critical coupling $g_{c}$
and independent of the number $m$ of the derivative of free energy.
Therefore, $A$ should be estimated consistently from the
various derivatives of the free
energy with respect to the magnetic field,
 and the scaling relation should be checked with  consistent
values of the exponents $\gamma_{st}$ and $\Delta$.

For n=1 Ising case, we have  an exact value of $A$ as 50.625. This value is
different from  the exact solution of two matrix model $A= 57.6 $[8,10]
due to our modification of the coupling constant $g$ by a factor
$(1-a^{2})^{2}$. We find that our series expansion also gives this critical
value
and by the  method of the extrapolation of the successive ratio of $c_{k}$
 for $k \rightarrow \infty$, we obtain quite accurately  the magnetic
field exponent
$\Delta=5/6$
from the series of $d^{2}F/d^{2}H$. Once the correct value of $A$ is obtained,
the exponents $\gamma_{st}$ and $\Delta$ are accurately estimated in our
problem, since the ratio of the coefficients is a monotonic function of the
order
of degree $k$.
The process of the analysis  may be shown
 by plotting $\Gamma= -(A_{c}- R_{k})k/A_{c}$ against $1/(k+1)$ where
$k$ is the order of the coupling constant $g$ and the ratio $c_{k}/c_{k-1}$
 is denoted by $R_{k}$.
This value $\Gamma$ approaches to the value of
$-3+\gamma_{st}+m\Delta$ for $k \rightarrow \infty$.
 In fig.1, the analysis of the free energy and the second derivative are
represented
 for n=1 Ising model.

 For n=2 Ising case, there appears a logarithmic singularity. Due to this
singularity, the exponent $\gamma_{st}$ becomes  underestimated
compared to the correct value $\gamma_{st}=0$.
However, the magnetic field exponent $\Delta$, which is  the difference between
two exponents
of the free energy and the second derivative of the free energy,  is not
influenced by this logarithmic behavior and it is
correctly estimated by the previous
ratio method explained before. We have used the values of  $a=1/4$ and
$A_{c}=54.0$ for $n=2$ Ising model.
 By the same method presented in  fig.1, we obtain $\Delta=0.75$ which
agrees with the  correct value.

  For n=4 (c=2) Ising model, we estimate as $A_{c}=59.2$, $a_{c}=0.23$
and $\Delta=0.79$. We have  the string susceptibility exponent
$\gamma_{st}=0.04$, which agrees with the
previous value [3,4].

  For n=6 (c=3) Ising model, we obtain $\Delta=0.81$ based on  the value
$a_{c}=0.21$ and $A_{c}=63.8$. The  exponent $\gamma_{st}
$ is estimated by this method as $\gamma_{st}=0.1$ which is consistent with
the previous result[3,4].
It may be interesting to note that the obtained
value of $\gamma_{st}$ of $c=3$ appears within the error bar of
 the value of pure
3d Ising specific exponent $\alpha$, which has been argued by various methods
as 0.11. This coincidence is surprising since our model for $c=3$ represents
the bosonic random surface behavior contrary to the fermionic random surface,
 which appears in the low temperature expansion of 3d Isind model.
If the sign factor due to the fermionic character has no important effect
and can be neglected, this coincidence may be understood [11].

  For n=8 (c=4) Ising model, we take $A_{c}=66.0$, $a_{c}=0.19$. We have
$\Delta=0.825$ and $\gamma_{st}=0.2$ for these values.

  The critival values of $A_{c}$ used here are consistent with the previous
  analysis [4]. The error bar may be $\pm 0.05$ for the estimated value of
$\Delta$. The accurate estimation of exponents for $n > 8$
becomes difficult within our
eighth order expansion, since the  fictitious or ghost  second peak
 of $\gamma_{st}$ appears for $ a > a_{c}$ as shown in the figures of
ref.[4]. There is a tendency that both $\gamma_{st}$ and $\Delta$ are
increasing with the increasing value of $c$.  It may be interesting to
find the value of $c$, at which the branched polymer behavior can be
seen.

 In the following, we discuss the  critical behavior near $a_{c}$
which appears in
the large system size, i.e. in the large k limit of the $g^{k}$ expansion.
  The coefficient $c_{k}$ of the free enegy in the large $k$ limit
becomes the partition function of n-Ising spin on a random surface, and we
denote the logarithm of $c_{k}$ devided by $k$ as $\tilde f$,
\begin{eqnarray}
  \tilde f & = & - \lim_{k\rightarrow \infty}( \ln c_{k})/k \nonumber \\
           & \sim & - \ln A(a,H)
\end{eqnarray}

This free energy $\tilde f$ has a singulality at $a_{c}$  and following
the conventional notations of the critical exponents of the specific heat
$\alpha$ and the susceptibility $\gamma$, we have

\begin{equation}
 \tilde f \sim (a - a_{c})^{2-\alpha}
\end{equation}
\begin{equation}
 \chi = d^{2}\tilde f/d^{2}H \sim (a-a_{c})^{-\gamma}
\end{equation}

The  cosmological constant $A(a,H=0)$  behaves near
the critical value of $a_{c}$ as

\begin{equation}
 A(a,0) = A(a_{c},0)+C(a-a_{c})^{1/\phi}
\end{equation}

\noindent
where $\phi$ is a crossover exponent and it describes the shift of the
cosmological constant due to the change of $a$. The crossover
occurs from the pure gravity behavior to n-Ising behavior at $a_{c}$.
The second derivative of the
free energy by $a$ becomes from (9) and (12),

\begin{equation}
 d^{2}\tilde f/d^{2}a \sim (a-a_{c})^{{1\over{\phi}}-2},
\end{equation}

\begin{equation}
 \alpha = 2 - {1\over{\phi}}
\end{equation}

 The exponent $\phi$ is represented  conventionally by $d\nu$, where $d$ is
Hausdorff dimension and $\nu$ is a critical exponent of the correlation
length. Since the specific heat exponent $\alpha$ is written conventionally
by $\alpha = 2 - d\nu$, we find

\begin{equation}
1/\phi = d\nu
\end{equation}

 The free energy $F$ has a scaling form for small H and $(a-a_{c})$,

\begin{equation}
 F = ( g - g_{c})^{2-\gamma_{st}}\psi({{a-a_{c}}\over{(g - g_{c})^{\phi}}},
{H\over{(g-g_{c})^{\Delta}}})
\end{equation}

Therefore, we have a scaling relation between two quantities,

\begin{equation}
 \vert a-a_{c} \vert^{1/\phi} \sim H^{1/\Delta}
\end{equation}

 and the magnetization $M$ and the susceptibility $\chi$ are expressed by

\begin{equation}
 M = d\tilde f/dH \sim H^{{1\over{\Delta}}-1}\sim (a_{c}-a)^{{1-\Delta}
\over{\phi}}
\end{equation}
\begin{equation}
\chi = d^{2}\tilde f/d^{2}H \sim
 H^{{1\over{\Delta}}-2}\sim (a-a_{c})^{{1-2\Delta}\over
{\phi}}
\end{equation}
\noindent
since we have $A(a,0)=A(a,H)+CH^{1/\Delta}$ and $\tilde f$ is given by (9).
Thus we find the expressions for the critical exponents $\alpha$,$\beta
$ and $\gamma$ in terms of $\Delta$ and $\phi$ which are valid
for $c > 1$ case:

\begin{equation}
 \alpha = 2 -{1\over{\phi}}, \beta = {{1-\Delta}\over{\phi}},
  \gamma = {{2\Delta -1}\over{\phi}}
\end{equation}

 From these equations, it is possible to compare our estimation of the
critical exponents with the  Monte Carlo simulations,
 especially
about $\beta/d\nu$ and $\gamma/d\nu$ since these quantities are described
by the magnetic field  exponent $\Delta$ from (15).
 We note that  the previous Monte Calro
results [5,7] for the exponents $\beta/d\nu$, $\gamma/d\nu$
 of $c=1$ does not agree with a correct result.

For $c \leq 1 $, we have $\beta=1/2$ and  $\Delta
=(3-\gamma_{st})/4$. Thus the exponent $\alpha$ becomes
$\alpha =  {2\gamma_{st}}/{(1+\gamma_{st})}$ [12].
This relation  does not hold for
$c > 1$ since we have already found that  the obtained value of  $\Delta$
 does not satisfy the relation
$ \Delta=(3-\gamma_{st})/4 $.

The magnetization $M=-d \ln A(a,H)/dH$ is obtained from the
calculation of the  inverse  cosmological constant $A(a,H)$ in our series
expansion up to order eight. Indeed we find that
the spontaneous magnetization
appears below $a_{c}$. However it is difficult to extract the
accurate value of the exponent
$\beta$ or $\gamma$
from this spontaneous magnetization. We have obtained rather
accurately the value
of $\beta/d\nu= 1-\Delta$ and $\gamma/d\nu= 2\Delta-1$
 through the study of $\Delta$,
 but the value of $d\nu$ remains unknown. If the spontaneous magnetization
is analyzed more precisely by the higher order expansion,
 it may be possible to
extract the value of $\phi=1/d\nu$.

 We have also tried the specific heat analysis for n=1 Ising model based upon
the series expansion up to order ten. From the coefficient $c_{k}$
of the series expansion about $g$ of the free energy
$F$ without a magnetic field, the ratio $R_{k}= c_{k}/c_{k-1}$ is
obtained. Then the quantity $A$ in (7) is estimated by a ratio method as
$A_{k}=(k+1)R_{k}-kR_{k-1}$. The second derivative of this quantity by
$a$ in (2) is proportional to the specific heat. For finite $k$, there
appears a maximum peak around a critical value $a_{c}=0.25$ as shown
in fig.2. The extrapolation of $A_{k}^{\prime\prime}$
 for $k\rightarrow \infty$
is simply done by plotting this value against $1/k$. As shown
in fig.2, we have a cusp  singularity for the
specific heat with an exponent $\alpha = -1$ and
a finite size scaling relation $A_{\infty}^{\prime\prime}-A_{k}^{\prime
\prime} \sim k^{
\alpha}$.  This method seems powerful to extract the exponent $\alpha =
2 - d\nu$, however for $c > 1$, the ghost singularity observed in
ref.[4] makes the analysis to be difficult without extending the series
expansion higher.

 As a summary, we have obtained numerically the value of the scaling
exponent of the magnetic field
$\Delta$ for $c > 1$ by the series analysis of n-Ising model,
 and we have shown how this  exponent is related to
other exponents.

 The author thank E. Br\'ezin, V. Kazakov and J. Zinn-Justin
for the valuable discussions on the present subject. He acknowledges
a hospitality of Ecole Normale Sup\'erieure. This work is supported by
the Grant-in-Aid for Scientific Research by the Ministry of
Education, Science and Culture.
\newpage

               {\bf  References}
\vspace{ 4mm }

\begin{description}
\item [{[1]}] E. Br\'ezin and V. A. Kazakov, Phys. Lett. B236 (1990) 144.\\
    D. J. Gross and A. A. Migdal, Phys. Rev. Lett. 64 (1990) 127.\\
    M. R. Douglas and S. H. Shenker, Nucl. Phys. B335 (1990) 635.
\item [{[2]}] V. G. Knizhnik, A. M. Polyakov and A. A. Zamolodchikov,
         Mod. Phys. Lett. A3 (1988) 819.\\
        F. David, Mod. Phys. Lett. A3 (1988) 651.\\
        J. Distler and H. Kawai, Nucl. Phys. B321 (1989) 509..
\item[{[3]}] E. Br\'ezin and S. Hikami, Phys. Lett. B283
         (1992) 203.
\item[{[4]}] S. Hikami and E. Br\'ezin, Phys. Lett. B295
         (1992) 209.
\item[{[5]}] C. A. Baillie and D. Johnston, Phys. Lett. B 286 (1992) 44.
\item[{[6]}] S. M. Catterall, J. B. Kogut and R. L. Renken, Phys. Lett.
             B292 (1992) 277.
\item[{[7]}] J. Ambj\o rn, B. Durhuus, T. J\'onsson and G. Thorleifsson,
             NBI-HE-92-35 (1992).
\item[{[8]}] D. V. Boulatov and V. A. Kazakov, Phys. Lett. B187 (1987) 379.
\item[{[9]}] B. Eynard and J. Zinn-Justin, Saclay preprint SPhT/93-029.
\item[{[10]}] M. L. Mehta, Commun. Math. Phys. 49 (1981) 327.
\item[{[11]}] J. Ambj\o rn, A. Sedrakyan and G. Thorleifsson, NBI-HE-92-80
              (1992).
\item[{[12]}] V. A. Kazakov, "fields, strings and critical phenomena",
             Les Houches 1988, edited by E. Br\'ezin and J. Zinn-Justin,
             P.369, North-Holland, Amsterdam (1990).
\end{description}

\newpage

      {\bf figure caption}
\begin{description}

\item[{fig.1}]   The dots show the values of
                 $\Gamma = -(A_{c}-R_{k})k/A_{c}$ plotted
                 against $1/(k+1)$, where $R_{k}$ are  ratios
                 of the coefficients
                 in the series of the free energy (upper line) of $n=1$ Ising
                 model
                 and of the second
                 derivative by a magnetic field (lower line).
                 The intersections of two lines with the vertical axis
                 shows the exponents $3-{\gamma_{st}}$ and
                  $3-{\gamma_{st}}-2\Delta$.
                  The value of $\Delta$ is obtained as 5/6.

\item[{fig.2}]   The second derivatives of the inverse
                 cosmological constant $A_{k}(a)$
                 by $a=\exp(-2\beta)$ are shown for $k=6,\dots,10$ in
                 the $n=1$ Ising model.
                 The broken line is the extrapolation for $k\rightarrow \infty$
                 and it shows a cusp singularity at $a=1/4$.

\end{description}

\end{document}